\documentclass[useAMS,usenatbib]{mn2e} 
\usepackage{aas_macros}
\usepackage{graphics}
\usepackage[pdftex]{graphicx}
\usepackage{epstopdf}
\usepackage{epsfig}  
\usepackage{natbib} 
\usepackage{dingbat}
\usepackage{float}
\usepackage{amsmath}
\usepackage{times}
\usepackage[varg]{txfonts}
\usepackage{verbatim} 
\usepackage{multirow,bigdelim} 
\usepackage{color}
\usepackage{vmargin}
\setmarginsrb{1.2cm}{1cm}{1.2cm}{1cm}{1cm}{1cm}{1cm}{1cm}

\newcommand{\hmsun}{{\, h^{-1}\rm~M}_\odot}

  \makeatletter
    \renewcommand{\paragraph}{\@startsection{paragraph}{4}{\z@}%
      {-3.25ex\@plus -1ex \@minus -.2ex}%
      {1.5ex \@plus .2ex}%
      {\normalfont\small\centering}}
     
    \renewcommand{\subparagraph}{\@startsection{subparagraph}{5}{\z@}%
      {-3.25ex\@plus -1ex \@minus -.2ex}%
      {1.5ex \@plus .2ex}%
      {\normalfont\small\centering}}
    \makeatother

\newcommand{\gadget}{{\sc Gadget}}

\newcommand{\ramses}{{\sc Ramses}}
\newcommand{\ramsescudaton}{{\sc Ramses-Cudaton}}

\newcommand{\hMpc}{{ \textit{h}$^{-1}$~Mpc}}

\title[LG reionization]{Reionization time of the Local Group and Local-Group-like halo pairs}
\author[Sorce et al.]
{Jenny G. Sorce$^{1,2}$\thanks{E-mail: \text{jenny.sorce@universite-paris-saclay.fr / jsorce@aip.de}}, 
Pierre Ocvirk$^{3}$, Dominique Aubert$^{3}$, Stefan Gottl\"ober$^2$, Paul R. Shapiro$^4$,
\and Taha Dawoodbhoy$^4$, Gustavo Yepes $^{5,6}$, Kyungjin Ahn$^{7}$, Ilian T. Iliev$^{8}$, Joseph S. W. Lewis$^{9,10}$\\ 
$^1$Universit\'e Paris-Saclay, CNRS, Institut d'Astrophysique Spatiale, 91405, Orsay, France\\
$^2$Leibniz-Institut f\"{u}r Astrophysik (AIP), An der Sternwarte 16, D-14482 Potsdam, Germany\\
$^3$Universit\'e de Strasbourg, CNRS, Observatoire astronomique de Strasbourg, CNRS UMR 7550, 11 rue de l'Universit\'e, 67000 Strasbourg, France\\
$^4$Department of Astronomy and Texas Cosmology Center, The University of Texas at Austin, Austin, TX 78712-1083, USA\\
$^{5}$Departamento de F\'{i}sica Te\'{o}rica, M\'{o}dulo 8  Universidad Aut\'{o}noma de Madrid, 28049 Madrid, Spain\\
$^{6}$Centro de Investigaci\'{o}n Avanzada en F\'{i}sica Fundamental (CIAFF), Universidad Aut\'{o}noma de Madrid, 28049 Madrid, Spain\\
$^{7}$Department of Earth Sciences, Chosun University, Gwangju 61452, Republic of Korea\\
$^{8}$Astronomy Center, Department of Physics \& Astronomy, Pevensey II Building, University of Sussex, Falmer, Brighton BN1 9QH, United Kingdom\\
$^{9}$Zentrum f\"ur Astronomie der Universit\"at Heidelberg, Institut f\"ur Theoretische Astrophysik, Albert-Ueberle-Stra\ss e 2, 69120 Heidelberg, Germany\\
$^{10}$Max-Planck-Institut f\"ur Astronomie, K\"onigstuhl 17, D-69117 Heidelberg, Germany\\
}

\begin{document}

\date{}

\pagerange{\pageref{firstpage}--\pageref{lastpage}} \pubyear{2022}

\maketitle

\label{firstpage}

\begin{abstract}
\indent 

Patchy cosmic reionization resulted in the ionizing UV background asynchronous rise across the Universe. The latter might have left imprints visible in present day observations. 
Several numerical simulation-based studies show correlations between reionization time and overdensities and object masses today. 
To remove the mass from the study, as it may not be the sole important parameter, this paper focuses solely on the properties of paired halos within the same mass range as the Milky Way.  
For this purpose, it uses CoDaII, a fully-coupled radiation hydrodynamics reionization simulation of the local Universe. This simulation holds a halo pair representing the Local Group, in addition to other pairs, sharing similar mass, mass ratio, distance separation and isolation criteria but in other environments, alongside isolated halos within the same mass range.
Investigations of the paired halo reionization histories reveal a wide diversity although always inside-out given our reionization model. 
Within this model, halos in a close pair tend to be reionized at the same time but being in a pair does not bring to an earlier time their mean reionization. 
The only significant trend is found between the total energy at $z=0$ of the pairs and their mean reionization time: pairs with the smallest total energy (bound) are reionized up to 50~Myr earlier than others (unbound). 
Above all, this study reveals the variety of reionization histories undergone by halo pairs similar to the Local Group, that of the Local Group being far from an average one. In our model, its reionization time is $\sim$625~Myr against 660$\pm$4~Myr (z$\sim$8.25 against 7.87$\pm$0.02) on average.

\end{abstract}


\section{Introduction}
 
 A considerable amount of effort is nowadays directed towards understanding the Epoch of Reionization (EoR), an era when radiation escaped from the first stars and galaxies, reached the neutral gas in the intergalactic medium and reionized it. In the last two decades, independent observations and measurements permitted determining a mean redshift for hydrogen reionization of z$\sim$8 \citep{2016A&A...594A..13P} and an end of reionization at z$\sim$6. This last redshift value has been obtained either through the Gunn-Peterson effect \citep{1965ApJ...142.1633G} in high redshift quasar spectra \citep[e.g.][]{2006AJ....132..117F,2011MNRAS.416L..70B,2015MNRAS.447..499M,2017A&A...601A..16B}, via the decline in observable Lyman-$\alpha$ emissions from high redshift galaxies  \citep[e.g.][]{2014MNRAS.440.3309D,2021ApJ...923...87C} with VLT- \citep[e.g.][]{2010MNRAS.408.1628S,2011ApJ...743..132P}, UKIRT- \citep[e.g.][]{2012MNRAS.422.1425C} and KECK- \citep[e.g.][]{2012ApJ...744..179S, 2013ApJ...775L..29T,2014ApJ...794....5T} based studies.\\
 
Since one of the first numerical work of  \citet{2001NewA....6..437G}, the theoretical side has not been left behind with the advent of cosmological simulations including sophisticated reionization models via numerical codes first without hydrodynamics \citep[e.g.][]{2006MNRAS.369.1625I,2007ApJ...671....1T} then intended for hydrodynamical simulations \citep[e.g.][]{2008MNRAS.387..295A,2009MNRAS.393.1090F} and subsequently included in hydrodynamical codes \citep[e.g.][]{2009MNRAS.396.1383P,2013MNRAS.436.2188R,2015MNRAS.454.1012A} giving rise to cosmological hydrodynamical simulated volumes including reionization \citep[e.g.][]{2016MNRAS.463.1462O,2020MNRAS.tmp.1438O,2021MNRAS.tmp.3440K}. These simulations cover the full range of scales to account for the patchiness \citep{2014ApJ...793..113P} of reionization  \citep[for reviews, see][]{2011ASL.....4..228T}. Indeed, the EoR is inhomogeneous on all scales from a few megaparsecs \citep[see Fig. 1 of][]{2020MNRAS.tmp.1438O} to tens of megaparsecs \citep{2014MNRAS.439..725I} and in order to correctly interpret current and upcoming observations, these inhomogeneities must be reproduced by simulations. All in all, the last decade witnessed a dramatic progress in our understanding of the timing of the EoR but its impact on galaxy formation is still not clearly identified. \\

For instance, star formation of low-mass galaxies might be suppressed by photoevaporation  \citep{2004MNRAS.348..753S,2005MNRAS.361..405I} and gas infall depletion \citep{1994ApJ...427...25S,2000ApJ...542..535G}. This mechanism constitutes a plausible scenario to the ``missing satellites problem''  \citep[e.g.][]{1999ApJ...522...82K} with the inhibition of star formation in low mass galaxies at early times \citep{2000ApJ...539..517B}. This scenario helps reproducing the satellite population of the Local Group \citep[e.g.][]{2009ApJ...696.2179K,2009MNRAS.400.1593M,2010ApJ...710..408B}. Observations seem to confirm that low-mass satellites of the Local Group have star formation histories compatible with an early suppression \citep[e.g.][]{2014ApJ...796...91B}. In that context, one can infer that the reionization history and time of objects may be tightly linked to their present day observations. It is then interesting to search for correlations between the small scale environment and even first and foremost the properties (e.g. mass ratio, distance separation, total energy) of the Local Group, a galaxy pair, and its reionization time.  Such correlations would emphasize the importance of comparing accurate modeling of the Local Group to its observed counterpart to pinpoint properly the reionization period and its effect still identifiable in today observations. For instance, \citet{2011MNRAS.417L..93O} showed that the properties of the Milky Way satellite population could retain, at z=0, some information about the timing and geometry of the ionizing UV radiation field during the EoR. \citet{2013ApJ...777...51O,2014ApJ...794...20O} further stressed that possibility with detailed studies based on the first constrained initial conditions produced by the CLUES consortium\footnote{https://www.clues-project.org} to actually reproduce the Local Group within its large scale environment. \\

Still previous studies already showed correlations between object masses and their reionization time \citep{2007MNRAS.381..367W,2018ApJ...856L..22A} as well as between densities and reionization time \citep{2018MNRAS.480.1740D}. 
Consequently, to emphasize the peculiarity of the Local Group reionization history with respect to other objects besides that linked to their mass, the latter must first be removed from the studies. We thus  select only objects of masses similar to that of the Milky Way. We go even further by investigating Local-Group-like pairs of halos including one replica, namely not only does this last pair share the same mass ratio and distance separation as the Local Group, but also the same large scale environment. We thus focus on the pairing effect and these pair property impacts on the reionization time as well as on where the Local Group reionization time stands with respect to the average\footnote{We note that \citet{2007MNRAS.381..367W} included a Local-Group-like sample in their study of reionization time as a function of the halo mass. Their selection criteria for Local-Group-like pairs are  quite similar to those we use and describe hereafter. However they neither investigate the impact of pair properties on the reionization time nor have a Local Group replica to compare with. Their sample size are also restricted compared to ours.}.\\

To that end, we use Cosmic Dawn (CoDa) II, a fully-coupled radiation hydrodynamics simulation of reionization of the local Universe whose general features were presented in \citet{2020MNRAS.tmp.1438O}. The initial conditions of such a simulation have been constrained with local observations to ensure that the simulated Large Scale Structure resembles that of the local Universe and includes a Local Group. The volume, (64~\hMpc)$^3$, of the simulation is large enough to include other galaxy pairs in different environments within a reasonable range of properties (i.e. in agreement with our knowledge of the Local Group in terms of isolation, distance separation and mass ratio, etc). With this simulation, we can then study in this paper the mean reionization time of the Local Group as well as that of other galaxy pairs to study potential correlations and compare these reionization times with those of isolated galaxies within the same mass range also present within the simulated volume.\\

 \begin{figure*}
\vspace{-0.8cm}
\includegraphics[width=0.8 \textwidth]{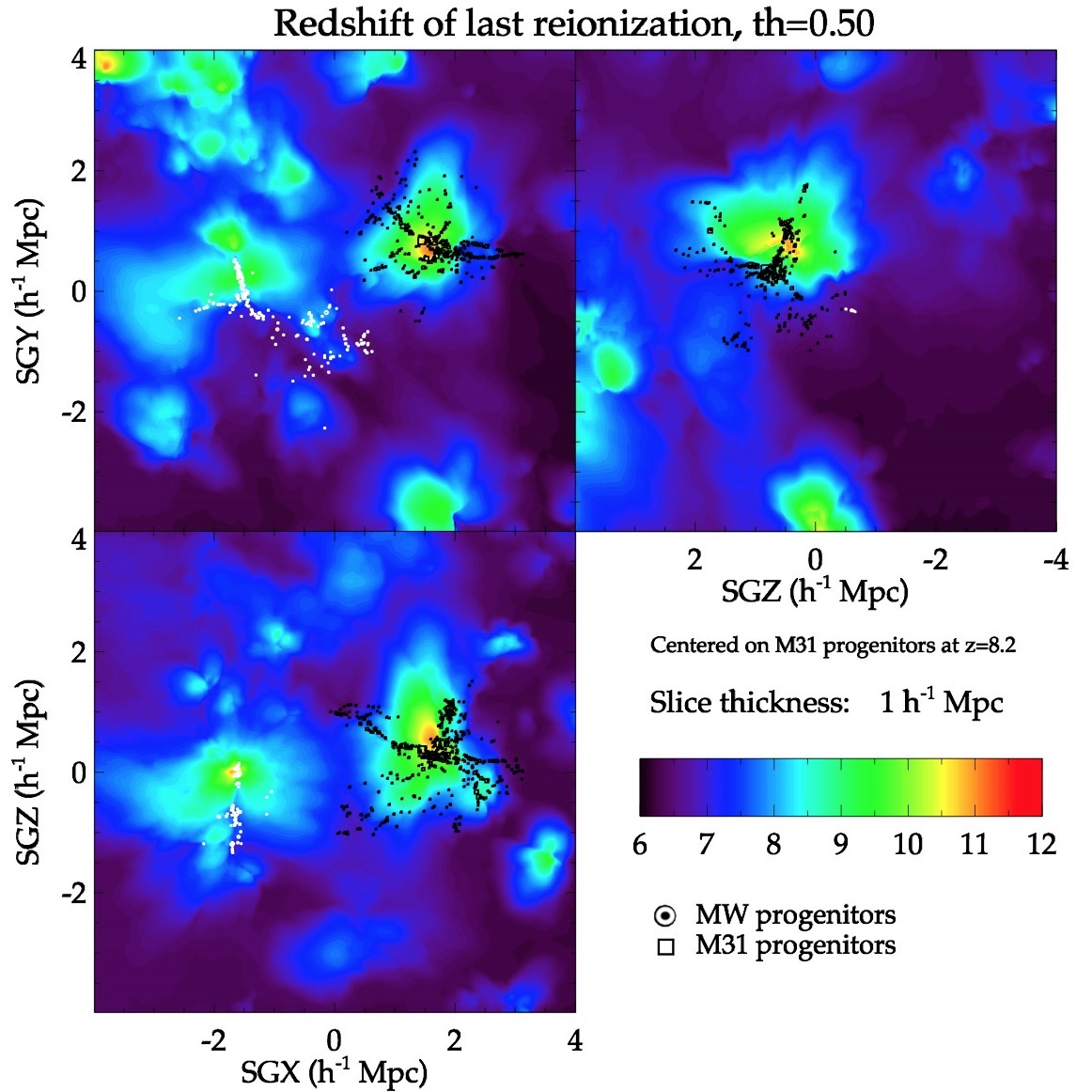}\\
\caption{XY, ZY and XZ supergalactic slices of the redshift of last reionization map of the Local Group progenitor. The redder the color is, the earlier the reionization of a given region on the map happened. On the contrary, the darker the color is, the later the reionization of the region happened. White circles stand for the progenitors of the Milky Way replica in a 1~\hMpc\ thick slice while the dark squares are the progenitors of M31 replica in that same slice. The slice is centered on the barycenter of M31's progenitors for a better visibility of the reionization of the latter. The progenitors have been identified at z=8.2 in agreement with the mean redshift for hydrogen reionization determined with Planck. Reionizations of M31 and the Milky Way appear to be inside-out.}
\label{fig:mwm31}
\end{figure*}

The second section of this paper describes the simulation from the building of its constrained initial conditions to its run through the reionization processes. It includes also the computation of the redshift of reionization maps. The third section introduces the selection of the Local Group simulacra - the replica (proper environment) and the look-alike (other environments) - as well as that of isolated halos within the same mass range. The fourth section investigates differences and correlations, if any, between isolated/paired halos and pair properties at z=0 and their mean reionization time before concluding. 
 

\section{The simulation}
 
A complete description of the simulation is available in \citet{2020MNRAS.tmp.1438O}. To summarize, the boxsize of CoDaII is 64~\hMpc, it contains 4096$^3$ particles and the same number of cells. The cosmology framework is that of Planck \citep[$\Omega_m$=0.307, $\Omega_\Lambda$=0.693, H$_0$=67.77~km~s$^{-1}$~Mpc$^{-1}$, $\sigma_8$~=~0.829,][]{2014A&A...571A..16P}. The Eulerian \ramses\ code \citep{2002A&A...385..337T} coupled with radiative transfer \citep[\ramsescudaton,][]{2016MNRAS.463.1462O} has been used. Since the simulation run was stopped at the end of the reionization (z=5.8) for computational reasons, a dark matter only simulation\footnote{This simulation, part of the MultiDark project, is available under the name ESMDPL-2048 at https://www.cosmosim.org/cms/files/simulation-data/} has been run in parallel with the same initial conditions but 2048$^3$ particles and with the Lagrangian  \gadget\ \citep{Springel2005} code to z=0. The two codes have been proven to perform identically at the same resolution in the dark matter case \citep{2016MNRAS.458.1096E}. This second simulation permits matching halos at z=0 with their progenitors in the dark matter simulation and by extension their Lagrangian regions during the reionization era in CoDaII.

\subsection{Constrained initial conditions}

Details of the method used to build the initial conditions constrained to resemble the local Universe thanks to local observational data are described in a set of papers \citep[e.g.][]{2015MNRAS.450.2644S,2016MNRAS.455.2078S,2018MNRAS.478.5199S}. These initial conditions are selected among a set of 200 initial conditions run at low resolution (512$^3$ particles, particle mass 1.7$\times$10$^8~\hmsun$) in the dark matter regime to first select that with the best Local Group at z=0. The Local Group belongs to the non-linear regime and is thus not directly constrained but induced by the local environment \citep{2016MNRAS.458..900C}. Investigating each simulation one at a time, pairs of halos at z=0 representing the Local Group in the vicinity of the center of the box (by construction) are retained if, at z=0:
\begin{enumerate}
\item their masses are between 5.5$\times$10$^{11}$ and 2$\times$10$^{12}~\hmsun$ masses\footnote{Masses used in this paper are derived with a friend of friend algorithm.},
\item there is no other halo more massive than 5.5$\times$10$^{11}~\hmsun$ within a sphere of radius 2~\hMpc,
\item their separation is smaller than 1.5~\hMpc,
\item their mass ratio is smaller than 2,
\item there are halos that could stand for M33 and Centaurus A (by far the most restrictive criterion),
\item they are located between 10 and 14~\hMpc\ away from the replica of Virgo.
\end{enumerate}
Left with a dozen halo pairs, we select the initial conditions giving birth to the pair (thus the simulation) that gathers both the requirements of a small separation and small mass ratio: at z=0, the distance between the two halos of the pair is 0.8~\hMpc\ and their mass ratio is 1.2 (masses are 1.5$\times$10$^{12}$ and 1.3$\times$10$^{12}~\hmsun$). 

The advantage of these selected new initial conditions with respect to those used for the first generation of CoDa \citep{2016MNRAS.463.1462O}, and thus any following studies of the Local Group reionization, is a Large Scale Structure that matches the local one down to the linear threshold ($\sim$3~\hMpc) on larger distances (the entirety of the box if not for the periodic boundary conditions) and with more accurate positions ($\sim$3-4~\hMpc) at z=0. In particular, this simulation contains a Virgo cluster at the proper position and with a mass in agreement with recent observational mass estimates \citep{2016MNRAS.460.2015S,2019MNRAS.486.3951S}. While the Virgo replica had a mass of 7$\times$10$^{13}~\hmsun$ in the first CoDa simulation, its mass is now 2.2$\times$10$^{14}~\hmsun$, i.e. the cluster is three times more massive. The boxsize is the only limiting factor to the cluster mass. Indeed, the longest modes in numerical simulations are important for the formation of the most massive objects \citep{2016MNRAS.460.2015S}. Therefore, it is expected that, in a small constrained box, the mass of the Virgo replica ends up being smaller than in a larger constrained box / that of the observed Virgo cluster. This simulation contains also other local objects \citep[cf. Fig. 2 of][]{2020MNRAS.tmp.1438O} such as a look-alike for Centaurus A and M33 (M$>$1$\times$10$^{12}~\hmsun$ and between 2 to 5~\hMpc\ from the Local Group pair). This is thus the first time that the reionization of the Local Group is studied within a large scale environment matching at this level the observed one.

\subsection{Reionization in CoDaII}
 
 As in the most simple typical hydrodynamical simulations, gas follows the ideal gas equation of state ($\gamma$=5/3) and star formation (efficiency of 0.02) occurs in cells which overdensity crosses a threshold ($\delta$=50). Additionally, the kinetic supernova (10\% of the stars with a 10 Myr lifetime) feedback implemented in \ramses\ \citep{2008A&A...477...79D} is turned on (10$^{51}$~erg/10~M$_\odot$). 

Regarding radiative transfer, each stellar particle is assumed to radiate during 10 Myr (massive stars lifetime) after that it becomes UV-dark. The radiation undergoes a mono-frequency treatment with an effective frequency of 20.28eV and an intrinsic emissivity of 4800 ionizing photons/Myr/stellar baryon. 

Finally, the speed of light is preserved thanks to a decoupling/coupling between GPUs and CPUs: while CPUs compute gas and gravitational dynamics, GPUs take care of the radiative transfer and ionization processus 80 times faster than required on CPUs.\\

In short, the resulting simulation accurately reproduces the gas properties and its interactions with ionizing radiation in the local Universe. In particular, the growth of typical butterfly-shaped ionized regions around the first stars and galaxies, together with photo-heating resulting in the progressive smoothing of the small scale gas structures, is visible. In addition, thanks to an improved calibration with respect to the first simulation of this type \citep{2016MNRAS.463.1462O}, CoDaII gives results in good agreement with most observable of the EoR: reionization history of the Universe as determined from the Lyman-$\alpha$ forest, cosmic star formation rate density, mean ionizing flux density, mean electron scattering optical depth seen by the Cosmic Microwave Background, UV luminosity function of galaxies down to an AB magnitude of -13 at 1600 \AA\ as well as over a broad range of magnitudes. In that respect, this simulation, although representative of only one reionization model, reproduces enough observational properties to warrant our study. Following results are however tied to this model. Given the originality of our study and the uniqueness of CoDaII, further comparisons cannot be conducted. Still conclusions of previous studies with a different focus, but in the same vein (mass / reionization time, density / reionization time correlations), are retrieved hereafter implying that our subsequent results are probably generalizable.

\subsection{Redshift of reionization maps}

After running the simulation, redshift of reionization maps can be built:
\begin{itemize}
\item For each timestep, the box resolution is coarsened to get cells of 0.06~\hMpc\ aside.  
\item One grid per timestep is then built with the same resolution. Each cell of a given grid is then filled with the mean ionized fraction (in percent) of the corresponding cell in the coarsened box at the same timestep. 
\item The threshold for a cell to be reionized is set to 50\%. Note that thresholds of 40 and 60\% yield the same results. Thresholds of 90 and 99\% only shift results to later redshifts but conclusions presented hereafter are unchanged. 
\item Taking each grid, from that corresponding to the earliest timestep to that standing for the latest, each cell of the map (again a grid with the same resolution) is filled with the redshift of the grid at which the corresponding cell value reaches the threshold for the first time. Maps of the first redshift of reionization are thus built. 
\item Maps of the redshift of last reionization are also derived: if, for a given snapshot, the cell value is again below the threshold, the redshift value in the corresponding cell of the map is reset until the threshold is reached again. Note that results presented in this paper are identical in both cases. All the numbers and plots in this paper are based on the latter. 
\end{itemize}

\begin{figure*}
\includegraphics[width=1. \textwidth]{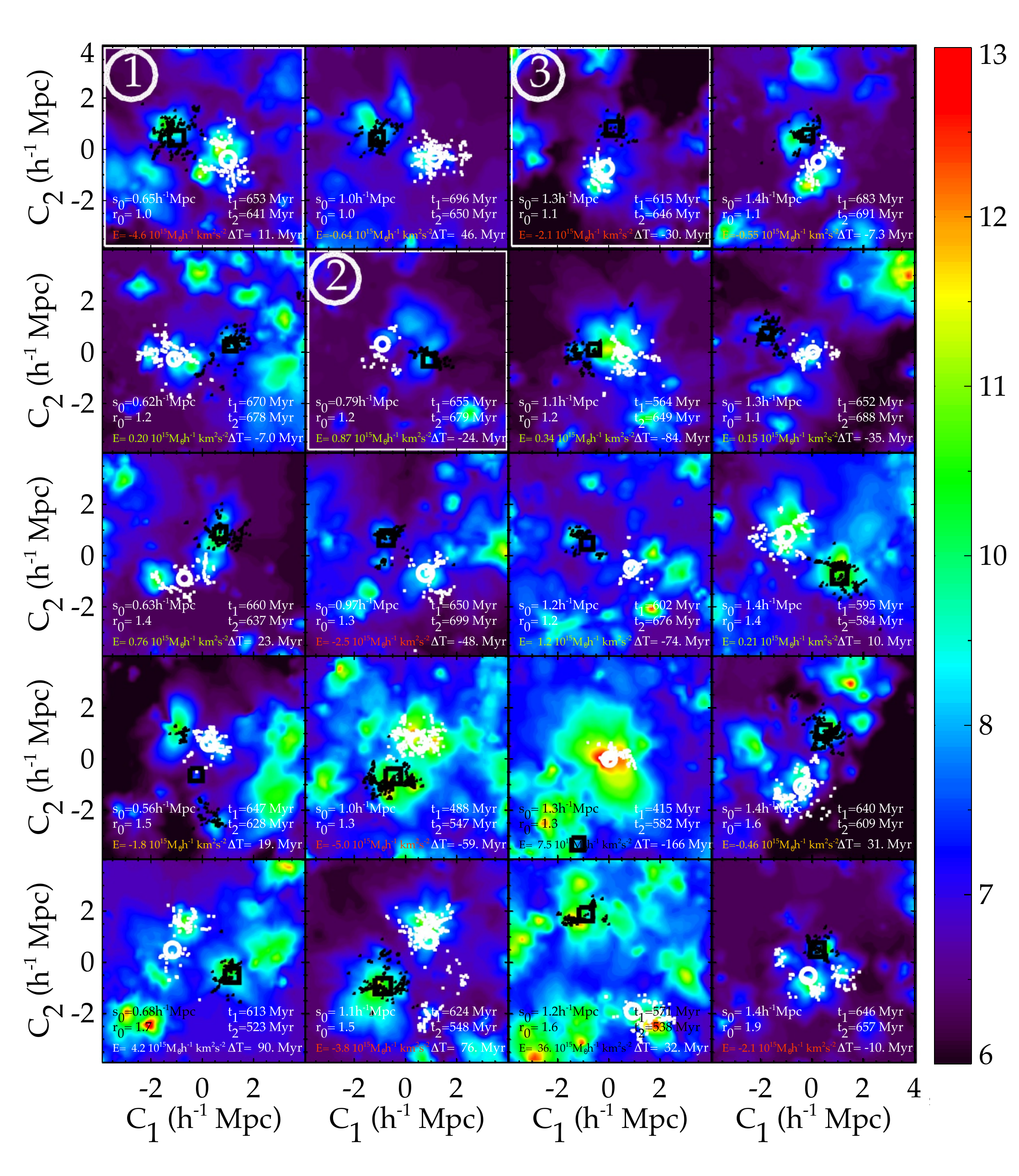}\\
\vspace{-0.5cm}
\caption{Slices of the redshift of last reionization map of several Local Group look-alike halo pairs. The redder the color is, the earlier the reionization of a given region on the map happened. On the contrary, the darker the color is, the later the reionization of the region happened. White small circles stand for the progenitors of the least massive halo of the pair in a 1~\hMpc\ thick slice while the small dark squares are the progenitors of the most massive halo of the pair in that same slice. The slice is centered on the barycenter of the pair and the slice is selected so as to give the best visibility of the global reionization of the pair. C$_1$ and C$_2$ are thus alternatively X, Y and Z coordinates. The progenitors are those identified at z=8.2. The largest white circles and black squares stand for the barycenter of the progenitors at z=8.2.  Pairs are ordered by increasing separation at $z=0$ (s$_0$)  from left to right and by increasing mass ratio at $z=0$ (r$_0$)  from top to bottom. The diversity of reionization histories is as wide as the diversity of pairs. t$_1$ and t$_2$ are the reionization time of the paired halos, most massive first, and $\Delta T=t_1-t_2$. Note that the third panel of the fourth row shows the most extreme case in the sense that it is the closest to the Virgo cluster hence it is reionized earlier but still inside-out. $E$ gives the total energy of the pair system. Embedded white numbers 1, 2, 3 are pairs referred to in the next figures.}
\label{fig:mapall}
\end{figure*}

\section{Local Group simulacra \& isolated MW/M31 mass halos}

\subsection{The Local Group replica}

Figure \ref{fig:mwm31} shows the redshift of last reionization maps centered on the Local Group progenitor replica whose position is determined thanks to the dark matter only simulation. The X, Y and Z supergalactic coordinates are defined similarly to the observational coordinates, with the Local Group replica at the center of the box at redshift zero and the box is oriented to find the main objects in the Large Scale Structure at the proper position. More precisely, the Local Group (Milky Way and M31) is identified in the 2048$^3$ dark matter only simulation. Subsequently the progenitors are identified in the different snapshots. Since the mean redshift for hydrogen reionization is about 8 \citep{2016A&A...594A..13P}, progenitors of the z=8.2 snapshot are overplotted on the map. Note that the differences between the positions of progenitors, if any, are insignificant for redshifts between 8 and 9. For a better visibility, only progenitors in a 1~\hMpc\ thick slice are shown. The XY, ZY and XZ slices are centered on the progenitors of M31 (chosen to be the most massive halo of the pair) to apprehend more distinctly the reionization of M31. Slices centered on the barycenter of the Local Group do not show as spectacularly the reionization patches. Equivalently, slices could be centered on the progenitors of the Milky Way. Since the visual is similar to that obtained for M31 except that the reionization patches are centered on the progenitors of the Milky Way, we prefer the quality to the quantity and focus only on the figure centered on M31, or more precisely on the barycenter of the progenitors of M31 at z=8.2. \\

The figure clearly shows that M31 reionizes itself (reionization inside-out)  given our reionization model. The reionization starts at quite an early redshift, about 11, where the density of progenitors is the largest and fades with the distance to the main aggregation of progenitors. The same behavior is observed for the Milky Way replica. A zoom out of the XY supergalactic slice of our redshift of last reionization maps is further shown in \citet{2020MNRAS.tmp.1438O} reinforcing the inside-out reionization of the Local Group conclusion even provided the Virgo cluster progenitor proximity as well as other large galaxy progenitors like Centaurus A and M81. Note that this zoom out is however centered on the barycenter of the Local Group rather than on for instance the barycenter of M31. Consequently, the reionization islands for both the Milky Way and M31 look smoother in \citet{2020MNRAS.tmp.1438O} than on Figure \ref{fig:mwm31}.

\subsection{The Local Group look-alike}

By construction, there is only one Local Group replica in the simulation, i.e. only one halo pair within the proper large scale environment. However, it is possible to select other halo pairs, with the same criteria as before minus Centaurus A, M33 and of course the Virgo cluster, in the dark matter only simulation at z=0. We retrieve 156 Local Group look-alike, i.e. not in the proper large scale environment but abiding by the  mentioned in section 2 criteria (i) to (iv), with (ii) less restrictive (radius decreased to 1.5~\hMpc).  Their halo progenitors are looked for in the dark matter only simulation at different redshifts. 
Figure \ref{fig:mapall} shows the redshift of last reionization maps centered on the barycenter of different pair progenitors selected by increasing separation at $z=0$ from left to right and increasing mass ratio at z=0 from top to bottom. Although these pairs present similar properties, the diversity of their reionization history is remarkable. Note, however, that all these pairs still seem to reionize themselves although at different time and pace within the context of our reionization model. The pair the closest to the Virgo cluster (third panel, fourth row) is no exception.

The reionization maps of three pairs, annotated 1, 2 and 3, are highlighted in the figure because they have respectively a reionization time similar for both halos (1), they have a separation and mass ratio at $z=0$ close to that of the Local Group replica (2), their mean reionization time is similar to that of the Local Group replica (3). They will be referred to in the next figures with these same numbers. 

\subsection{MW-M31 mass halos}

Some theoretical studies of the reionization of the Milky Way use isolated halos. It is worth checking the relevance of this choice by comparing the reionization time of isolated halos to that of paired halos within the same mass range. We select halos abiding by criteria (i) and (ii) with the additional enforcement that they actually have to be the only halo more massive than 5$\times$10$^{11}~\hmsun$ within 1.5~\hMpc. We find 1082 such halos. Our sole interest here resides in comparing the reionization time of these isolated halos to those in pairs to check the influence or not of being in a pair. Reionization times are derived in the next subsection.

\subsection{Mean redshift of reionization}

To compute the mean redshift of reionization of the Local Group simulacrum (replica and look-alike) progenitors as well as that of isolated halo progenitors within the same mass range at $z=0$, we proceed as follows for each halo (either isolated or of the pair): 
\begin{itemize}
\item Halo progenitors at $z_1$ are superimposed to the redshift of last reionization map. 
\item For each halo progenitor, the redshift value in the cell it falls in is recorded if it is $z_1$ or earlier. 
\item The procedure is repeated with halo progenitors at $z_2<z_1$. 
\item  For each halo progenitor, the redshift value in the cell it falls in is recorded if it is $z_2 $ or earlier and $z_1$ has not been attributed to the halo progenitor via its own halo progenitors.
\item The procedure is repeated with halo progenitors until $z=6$ (from $z=20$). For each halo progenitor, the redshift value in the cell it falls in is recorded if it is $z_i$ or earlier and no redshift of reionization has yet been attributed to the halo progenitor via its own halo progenitors. 
\item The weighted-by-progenitor-mass mean of the redshift values obtained for all the progenitors of one halo of the pair is computed. 
\end{itemize}

Note that the medians were also computed but results presented in the next section are identical. If the maximum rather than the mean is used, then the plots are simply shifted in redshift by about 1. If only the most massive of the halo progenitors is considered then values are approximately shifted in redshift by about 4. If all the dark matter particles are considered instead of progenitors then the mean reionization time is delayed in agreement with \citet{2018ApJ...856L..22A} findings. In the following for the reionization time of pairs, deriving their mean reionization redshift rather than their mean reionization time does not change the results obtained given our reionization model.

\begin{figure*}
\includegraphics[width=0.49 \textwidth]{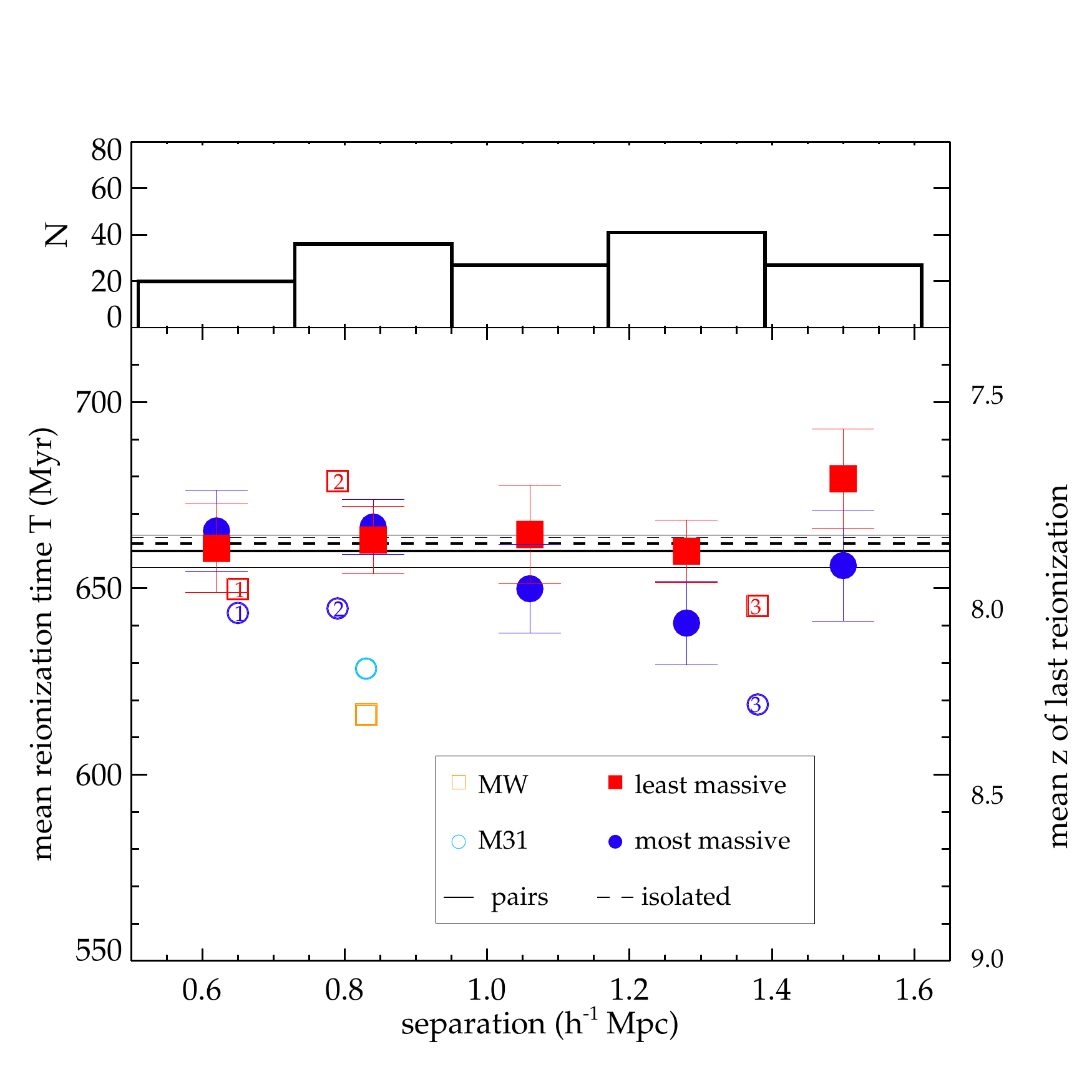}\hspace{-0.36cm}
\includegraphics[width=0.49 \textwidth]{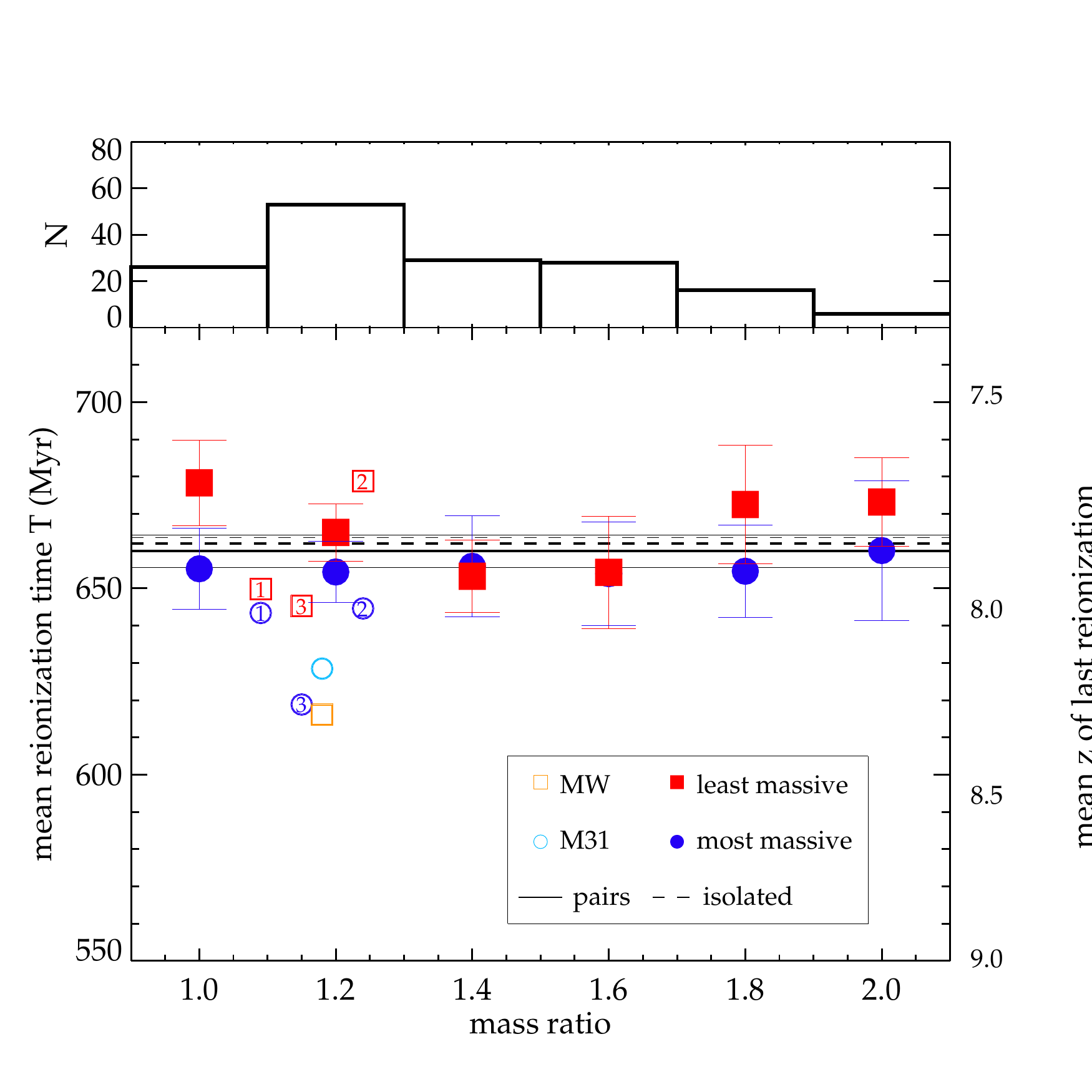}
\vspace{-1cm}

\includegraphics[width=0.49 \textwidth]{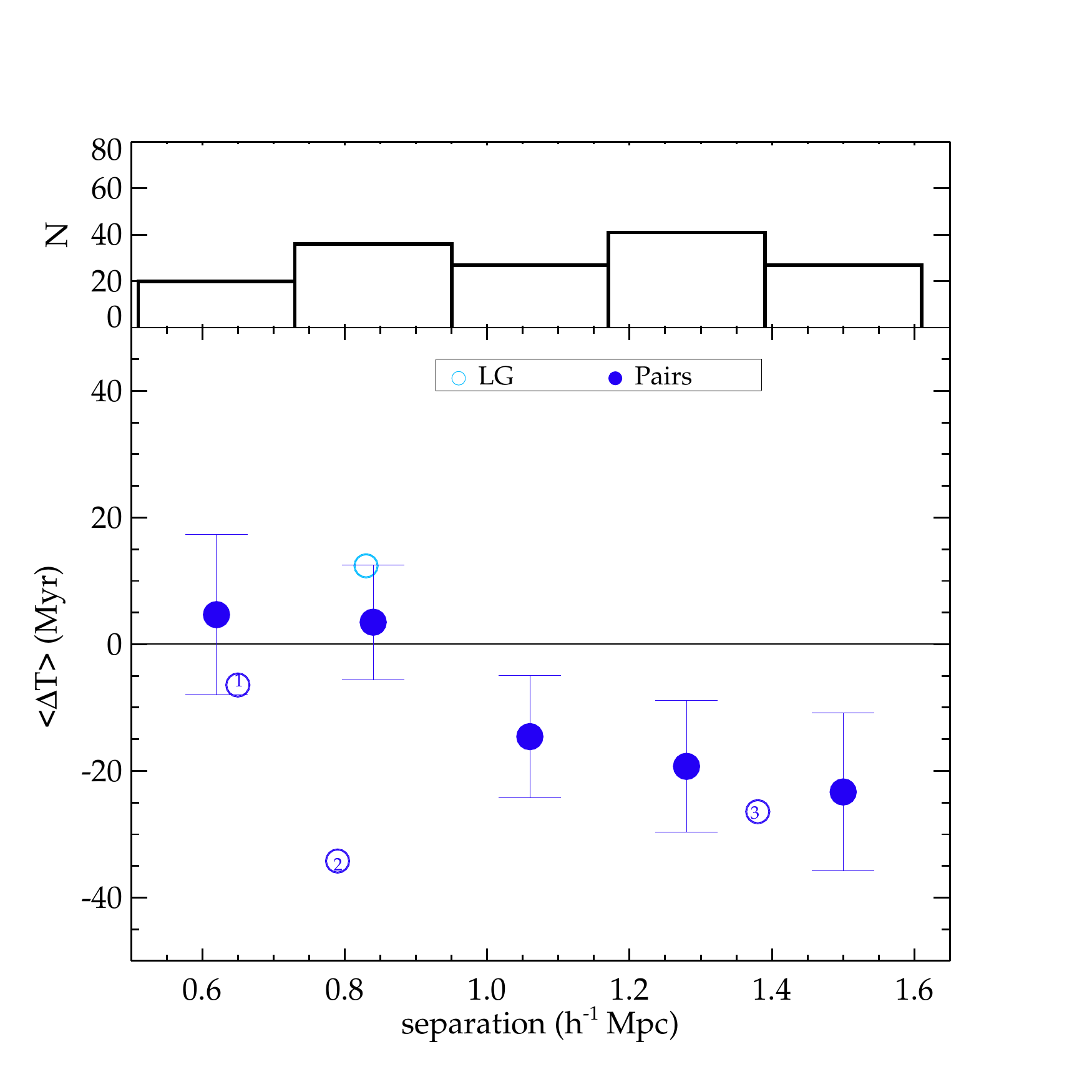}\hspace{-0.36cm}
\includegraphics[width=0.49 \textwidth]{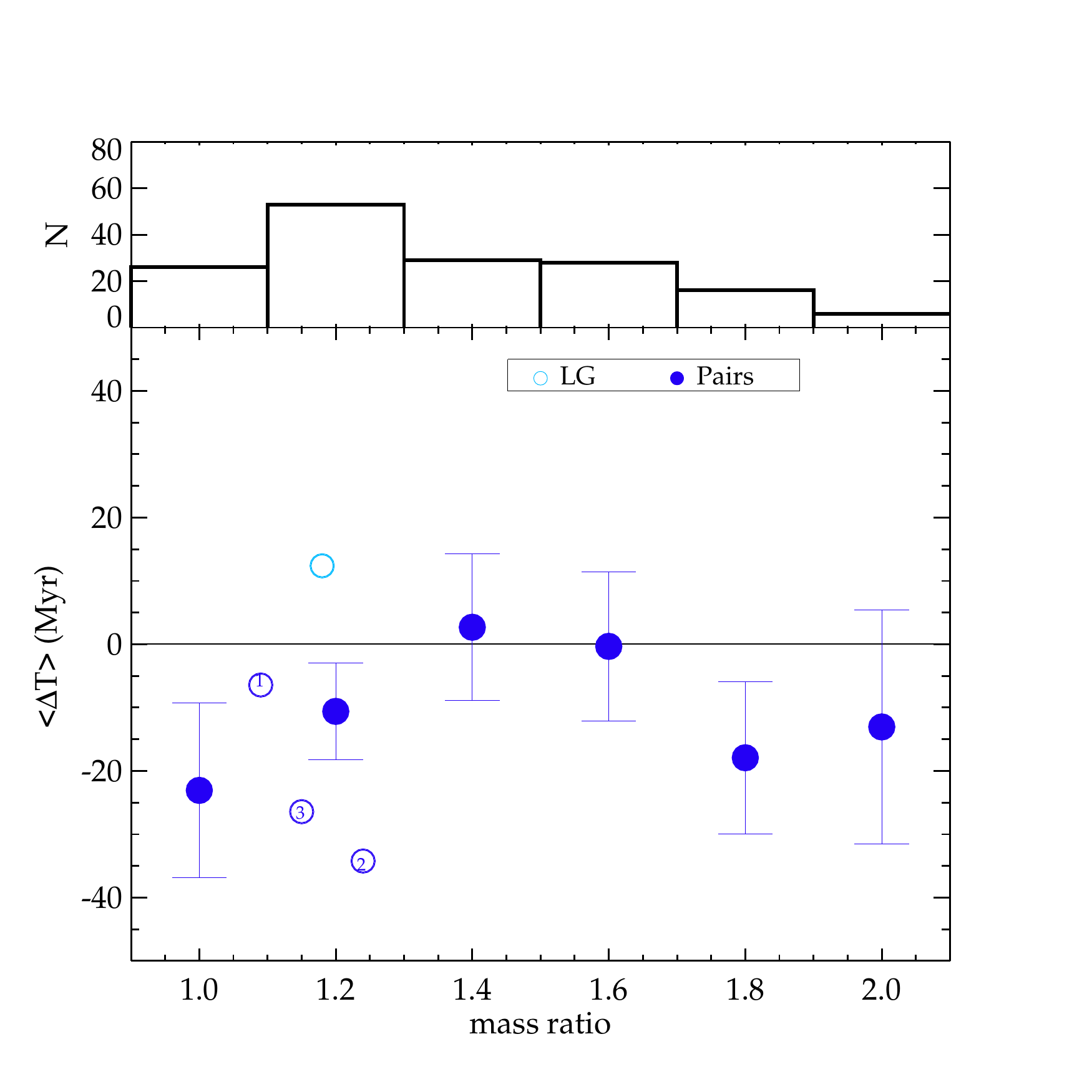}
\vspace{-0.5cm}
\caption{Top: Mean redshift of last reionization of Local Group simulacra (replica: orange square and light blue circle, look-alike: filled red square and filled blue circle) vs. their separation (left) and mass ratio (right). Bottom: Difference between reionization times of the two halos of a pair -- reionization time of the most massive halo minus that of the least massive halo of the pair (replica: light blue circle, look-alike: filled blue circle) vs.  their separation (left) and mass ratio (right). Error bars stand for the error on the mean times (top) and their mean differences (bottom). The circled numbers, 1, 2 and 3, stand for the pairs highlighted with the same numbers in Fig. \ref{fig:mapall}. The upper quadrant in each panel shows the distribution of halo pairs in the different bins.}
\label{fig:ratiodist}
\end{figure*}

\begin{figure*}
\includegraphics[width=0.49 \textwidth]{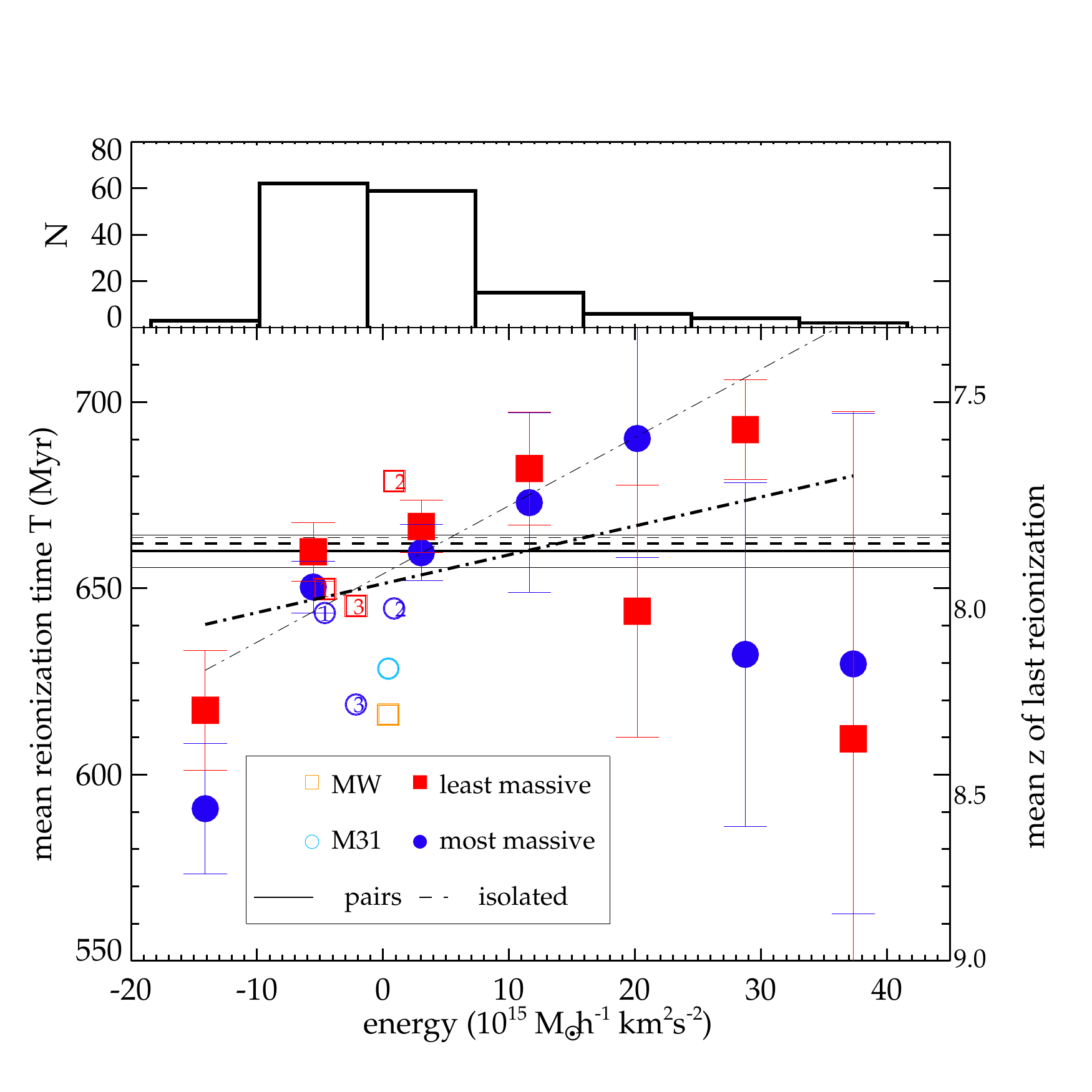}\hspace{-0.36cm}
\includegraphics[width=0.49 \textwidth]{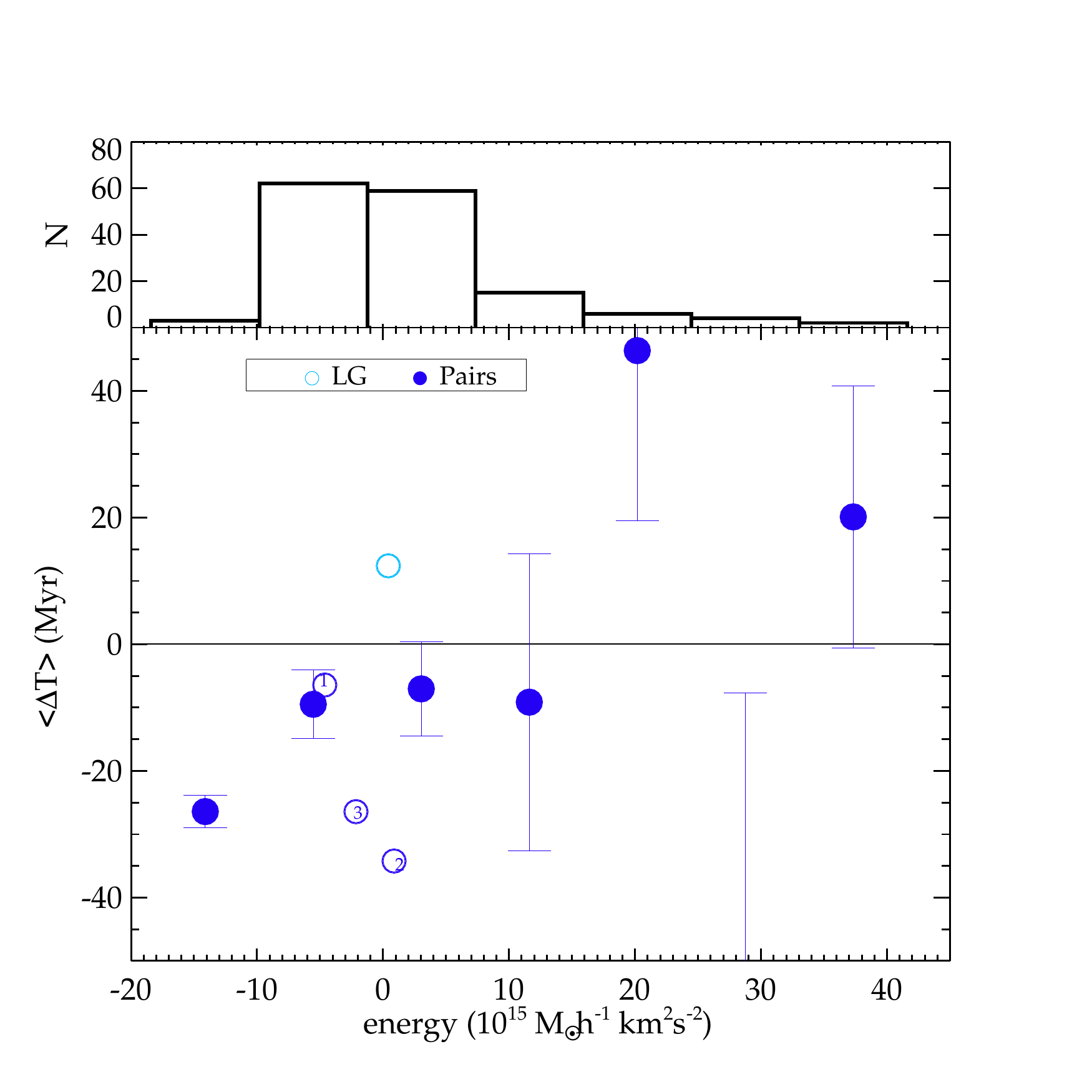}
\vspace{-0.5cm}

\caption{Same as Figure \ref{fig:ratiodist} but for the total energy of the pairs. The additional dot-dashed thick and thin black lines in the left panel are first degree polynomial fits to all the binned points and to the five first binned points respectively.}
\label{fig:energy}
\end{figure*}

 \section{Reionization time \& correlation}
 
With the mean reionization time for all the isolated halos and halo pairs including the Local Group replica, the potential differences and correlations between the mean reionization time of isolated and paired halos and the properties of the latter at z=0 (e.g. mass ratio, separation and total energy) can be studied. Consequently, all the following results are valid within the context of our reionization model but can most likely be extended, given the remarkable agreement between our simulation and observations of the EoR as well as with previous studies using other simulations of that epoch, as explained in the introduction.

\subsection{Paired vs. isolated halos}  
The mean reionization time of isolated halos within the same mass range as the Milky Way and M31 is 662$\pm$2 Myr (z=7.85$\pm$0.02). The mean reionization time of the Local-Group-like pairs is 660$\pm$4~Myr (z=7.87$\pm$0.04). Isolated and paired halos are thus reionized on average at the same time as shown by the thick dashed and solid lines on Figure \ref{fig:ratiodist}. The thin dashed and solid lines are the error on this mean. 
Consequently, the correlation between the reionization time and the halo mass \citep{2018ApJ...856L..22A} overcomes the correlation between densities and reionization time \citep{2018MNRAS.480.1740D} at these scales. Indeed,  \citet{2018MNRAS.480.1740D} use coarser grids than \citet{2018ApJ...856L..22A} to estimate the mean reionization time while we use intermediate grids, this implies that at the halo level, halos of the same mass are reionized at the same time, still if they are in groups they reionize their environment faster thus at an earlier time. \\

To study the dependence on separation (left panel of Figure \ref{fig:ratiodist}) and mass ratio (right panel of Figure \ref{fig:ratiodist}) of Local-Group-like pairs at z = 0, we classify them by bins (top quadrant of each panel). Zero and first degree polynomials can be fitted to the mean reionization time of both halos in the pair. Weights for each points in the fit are then the square root of the sum in quadrature of the error on the means of both halos' reionization times. 

\subsection{Separation at z=0} 

There is no clear trend between the mean reionization time and the separation of the two halos in the pair at $z=0$. The zero degree polynomial is completely consistent with the points.\\

However, there is a  slight trend suggesting that close by halos in a pair at $z=0$ tend to have the same reionization time. Conversely, when halos are further apart, the most massive halo of the pair is on average reionized earlier than the smallest one. This last observation of the most massive halos being reionized earlier than the least massive ones is in agreement with results of \citet{2018ApJ...856L..22A}. It is thus interesting that this trend is not preserved when halos of masses such as that of the Milky Way are in close pair. It seems that when they are separated by more than 1~\hMpc\ they cannot be considered on average as the same type of pairs as those closer than 1~\hMpc. This is confirmed by the bottom left panel of the same figure: close by halos in a pair at $z=0$ have been reionized at a similar time. For further apart halos in a pair, the most massive halo of the pair is on average reionized earlier. \\

The overplotted replica of the Local Group (opened blue circle and orange square) shows that it presents the same trend as the other paired halos on average given its separation: the Milky Way, chosen to be the least massive halo of the pair, is reionized at the same time as M31, even slightly earlier. However the Local Group pair is not an average pair since it is reionized earlier by $\sim$40~Myr than predicted on average. Note that the opposite conclusion can be reached with the pair number (2) that shares the same distance separation and mass ratio as the Local Group: although it has a mean reionization time in agreement with the average one,  the most massive halo of the pair is reionized earlier. This highlights again the diversity and complexity of reionization history.\\

\subsection{Mass ratio at z=0} 

Again there is no clear trend between the mean reionization time and the mass ratio of the two halos in the pair at $z=0$. The zero degree polynomial is completely consistent with the points.  \\

There is always a slight tendency for the least massive halo of the pair to be reionized later, but at intermediate mass ratios, as shown on the right bottom panel of the figure. This is again in agreement with the fact that the smallest halos are reionized later \citep{2018ApJ...856L..22A}. \\

Note that the minimum difference between reionization time can be understood as a selection bias effect: because of the restricted mass range, the least massive halos of pairs that have a mass ratio close to 2 are on average smaller than those of pairs with intermediate mass ratio. Consequently, they tend to be reionized later.  

More interestingly, the Local Group replica, yet again, is reionized earlier than other similar pairs on average given its mass ratio. The pair number (3), that shares the same reionization time as the Local Group, has the same mass ratio but the separation between its halos is almost twice that of the distance between the Milky-Way and M31 that is quite constrained from observational measurements. Accordingly, in agreement with the previous subsection, unlike for the Local Group replica, the most massive halo is reionized earlier than the least massive one (bottom right panel).

 \subsection{Total Energy at z=0} 
 
There is thus no correlations between separation and mass ratio at z=0  and the mean reionization time of the pairs. This is most probably due to the fact that these properties have evolved since reionization. The slight trend indicating that paired halos seem to be reionized at the same time when they are closer than further apart calls for finding a property that is more preserved than the separation through cosmic time. The known mass - reionization time correlation found in a different EoR simulation \citep{2018ApJ...856L..22A}, also hinted at by our above results, invites for the total mass to be included in this indicator as well. 

We choose the total energy of the system and consider the pairs as isolated systems. The sum of their kinetic and potential energies should thus be preserved through cosmic time. The kinetic energy relies on both the mass of the system and the relative velocity of the two halos. Mass and separation of the halos are part of the potential energy. \\

Figure \ref{fig:energy} shows the clear correlation between the total energy of the pairs at $z=0$ and their mean reionization time especially for the smallest total energy. The dot-dashed thin and thicker lines show the first degree polynomial fits for all the binned datapoints and restricted to the five first binned datapoints respectively. Namely tightly bounded system tend to be reionized on average earlier than those poorly connected. It appears as if they constitute  one single halo, thus are on average more massive than when considered separately and by extent are reionized earlier because of the mass - reonization time correlation \citep{2018ApJ...856L..22A}. The right panel shows however that there is no correlation between the total energy of the pair and the difference between the reionization times of the halos of a same pair. It might be hinted though that the difference between reionization times of paired halos varies less from one tightly bounded pair to the other than for other pairs: the huge variance of the difference in the latter case implies  that reionization times can differ quite a lot between halos of pair with high energy. In that respect, the sign of the total energy permits distinguishing at z=0 systems of two halos that evolved concomitantly, thus that were reionized at most within $\sim$30Myr apart, against others constituted of two halos that might have been reionized at quite different times. \\

 \begin{figure*}
\includegraphics[width=0.49 \textwidth]{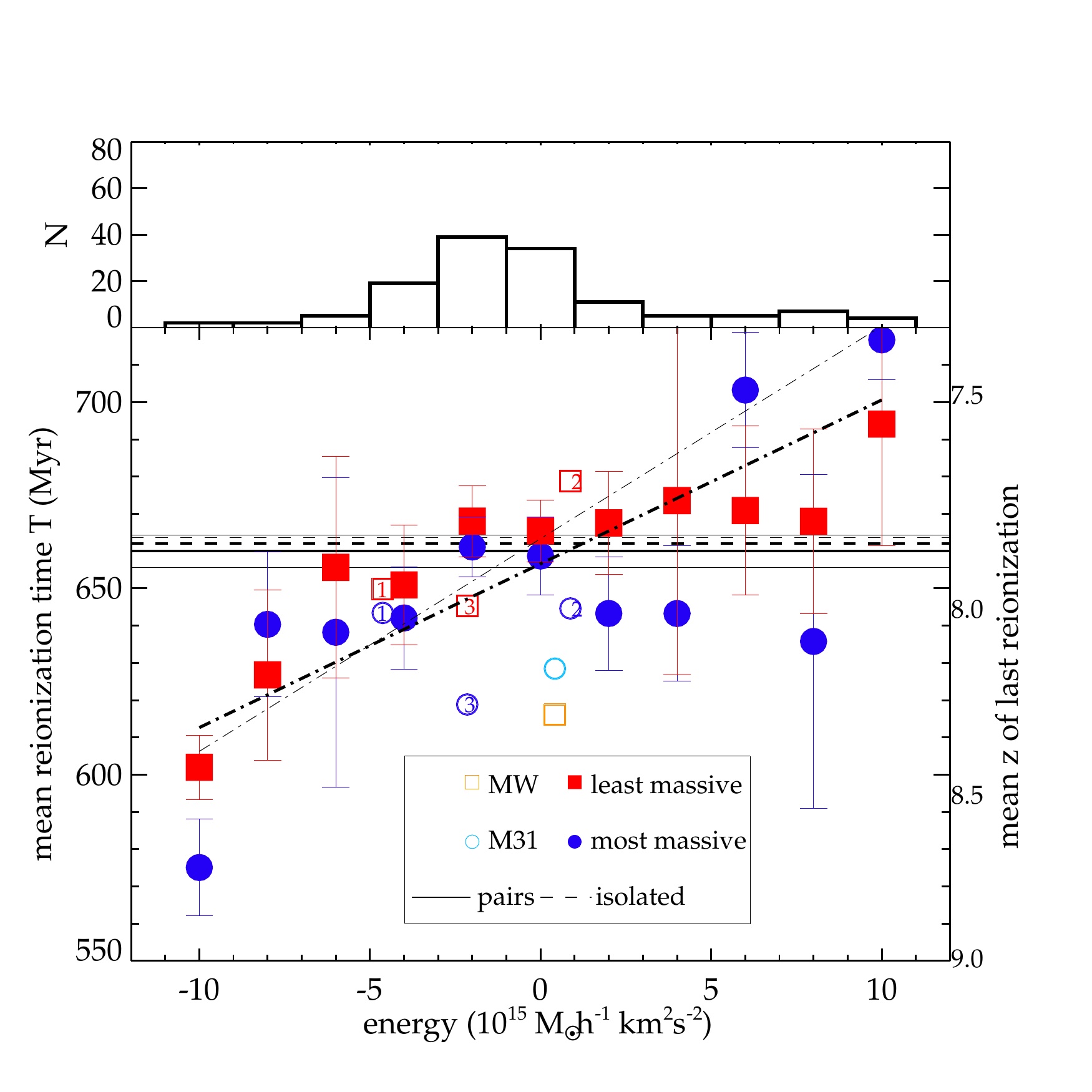}\hspace{-0.36cm}
\includegraphics[width=0.49 \textwidth]{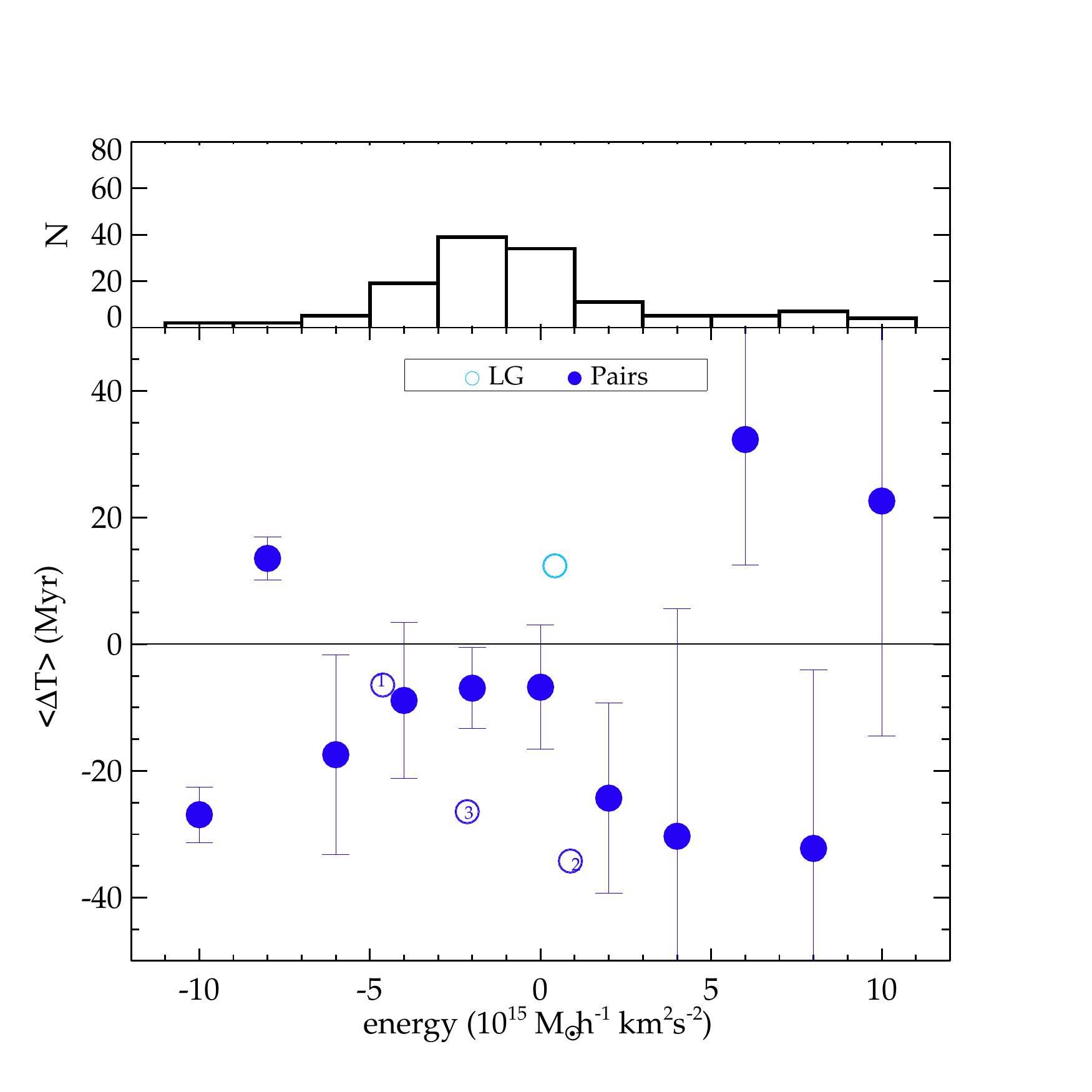}
\vspace{-0.5cm}

\caption{Same as Figure \ref{fig:energy} but zoomed on the [-10,10]~M$_\odot$h$^{-1}$km$^2$s$^{-2}$ total energy range. The dot-dashed thick and thin black lines in the left panel are first degree polynomial fits to all the binned points and to the binned points minus the last five bins respectively.}
\label{fig:energybis}
\end{figure*}

A deeper study might reveal that bound pairs are within islands of islands of reionization favoring radiative cross-talks and mutual reionization between halos while the other pairs are in simple islands. This appears quite clearly on Figure \ref{fig:mapall} for the pair number (1). This pair has a negative total energy and can thus be defined as bound. Its reionization map shows that each halo\footnote{More accurately speaking progenitors of each halo.} of the pair is within its reionization bubble (green color) and that the reionization patches of the two halos are within a global reionization bubble with a bridge between the two halos of the pair (light blue color embedding the two green patches on a violet background). On the contrary, the pair in the third column of the last row, with a high positive total energy, presents islands for each halo but these islands are not embedded into an island isolating them from the background and permitting privileged cross-talks between each other (light blue color filling the full background). It is also true for the pair in the first column of the same row that is a less extreme case (reionization time more similar to the pair number (1)). Pairs of intermediate energy like the pair number (2), or its left neighbor on the figure, or the Local Group replica on Figure \ref{fig:mwm31}, depicts an intermediate scenario: slight hints of islands in islands but not fully isolated from the background/other islands. In future work, it might be interesting to further study this qualitative argument with structure finders following \citet{2021arXiv211111910T}.

Because most of the pairs are within the intermediate energy range, we further refined the bins for intermediate energy values in Figure \ref{fig:energybis}. The correlation holds: the more bound a system is (negative total energy), the earlier it is reionized on average.\\

Despite this correlation, the Local Group replica is still quite off the mean suggesting that although the total energy of the system is a great indicator for the mean reionization time of a pair, the large scale environment has also an impact. This is suggested by the correlation reionization time - overdensity  \citep{2018MNRAS.480.1740D} but also by preliminary indicators regarding the neighbors of the pairs. For instance, the pair closest to the Virgo cluster indeed is reionized the earliest (see third panel of the fourth row in Figure \ref{fig:mapall}. However, the focus of this paper is solely on the intrinsic properties of the pairs. In addition, a statistical study would be difficult to conduct given our boxsize and the fact that the simulation reproduces the local Universe: indeed there is by definition only one massive halo (Virgo), the others being outside (like Coma) or impacted by boundary conditions (like Centaurus). In any case, one can note that the reionization still appears to be inside-out, given our reionization model, even when pairs are close to the Virgo counterpart.


\section{Conclusion}

Within the past two decades, the EoR understanding has been improved. However the impact of reionization on galaxy formation remains to be fully understood on many aspects. From patchy cosmic reionization, the ionizing UV background did not rise simultaneously across the Universe thus affecting galaxy formation in diverse ways. This asynchrony possibly left an imprint, still observable today, for instance in the properties of galactic satellite populations. Previous studies, based on different EoR simulations,  showed correlations between halo mass and reionization time as well as between overdensity and reionization time. It is thus interesting to go further determining whether there is a correlation between reionization time and the intrinsic properties of similar mass galaxies either isolated or in pairs. To give a two-fold goal to our project, we choose the Local Group, the Milky Way - M31 pair of galaxies, as the mass reference and our object of study.  By selecting both isolated and paired halos of the same mass as those of the Local Group, the mass - reionization time dependence is removed from the analyses. We are thus able to determine how typical the reionization history of the Local Group is compared to other pairs within our EoR model. \\

Our implemented reionization model applied to constrained local Universe initial conditions gave rise to a simulation of the reionization of the local Volume, named CoDaII \citep{2020MNRAS.tmp.1438O}, within which we can identify a Local Group replica (proper environment) and Local Group look-alike halos (other environment but similar mass ratio, isolation criterion and distance separation) as well as isolated halos within the same mass range. We study the differences and correlations between the redshift of last reionization of isolated and paired halos' progenitors and the intrinsic properties of the halos at $z=0$ (mass ratio, distance separation and total energy). Since the reionization simulation ends at z=5.8, a dark matter only simulation counterpart is used to determine properties of halos at z=0 and to identify their progenitors.\\

At fixed mass, our findings, given our reionization model, are described hereafter. Note that they are most probably generalizable since we showed in \citet{2020MNRAS.tmp.1438O} excellent agreements with observational data of the EoR and, we recover in this paper hints of similar results (like the reionization time - mass and reionization time - overdensity correlations) to those of previous studies of other EoR simulations.
\begin{enumerate}
\item the reionization histories of halo pairs are diverse.
\item whatever the configuration of the halo pair is, they appear to be reionized inside-out.
\item isolated halos tend to be reionized on average at the same time as paired halos: 662$\pm$2~Myr vs. 660$\pm$4~Myr, i.e. z=7.85$\pm$0.02 vs.  z=7.87$\pm$0.04 on average.
\item the distance between the halos of the pair influences the reionization time of the least massive halo of the pair: although, as expected, the most massive halo is reionized about 20-30~Myr earlier than the least massive halo on average, when halos are closer than 1~\hMpc, both halos are reionized at the same time. This suggests the existence of some amount of radiative cross-talk and mutual reionization between the progenitors of the paired halos during the EoR.
\item the mass ratio and the separation of the halo pair are overall not correlated with their mean reionization time.  
\item the indication that paired halos tend to be reionized at the same time when in close proximity suggest that $z=0$ separation and mass ratio do not retain a sufficient amount of earlier configurations. 
\item assuming the pairs to be isolated systems, their total energy is much better preserved across cosmic time. Thus a strong correlation with the mean reionization time is found: the smaller the total energy is, the more bounded the pair is and the earlier the reionization time of the pair is (by up to$\sim$50~Myr). 
\item still the Local Group does not have a reionization time that matches average predictions suggesting that the (large scale) environment is another important key to get the Local Group reionization history properly: 625~Myr vs. 660$\pm$4~Myr, i.e. z=8.25 vs. z=7.87$\pm$0.04 on average.
\end{enumerate}

Above all, this study reveals the variety of complex reionization histories that halo pairs such as the Local Group may have undergone. In addition, the reionization history of the Local Group is far from being that of an average pair of halos within the same mass range. It implies the necessity to have a proper pair of halos with appropriate intrinsic properties (in particular the total energy linked to the mass and separation of the two halos), preserved with cosmic time, in the right environment (in particular the presence of close-by massive neighbors), to study its reionization history and properly identify its effect that may still be identifiable in today's observations of the Local Group. 

\section*{Acknowledgements} 

JS acknowledges support from the French Agence Nationale de la Recherche for the LOCALIZATION project under grant agreements ANR-21-CE31-0019. 
PRS was supported in part by US NSF grant AST-1009799, NASA grant NNX11AE09G, NASA/JPL grant RSA Nos. 1492788 and 1515294, and supercomputer resources from NSF XSEDE grant TG-AST090005 and the Texas Advanced Computing Center (TACC) at the University of Texas at Austin. 
TD is supported by the National Science Foundation Graduate Research Fellowship Program under Grant No. DGE-1610403.
GY acknowledges financial support from  MICIU/FEDER under project grant PGC2018-094975-C21.  
KA is supported by NRF-2016R1D1A1B04935414, 2021R1A2C1095136, and 2016R1A5A1013277. KA also appreciates the APCTP and the KASI for their hospitality during completion of this work.
ITI was supported by the Science and Technology Facilities Council [grant number ST/L000652/1]. 
This research used resources of the Oak Ridge Leadership Computing Facility at the Oak Ridge National Laboratory, which is supported by the Office of Science of the U.S. Department of Energy under Contract No. DE-AC05-00OR22725. CoDa II was performed on Titan at OLCF under DOE INCITE 2016 award to Project AST031.
The ESMDPL simulation has been performed on the GCS Supercomputer SuperMUC-NG at the Leibniz Supercomputing Centre (www.lrz.de) within the project pr74no funded by the Gauss Centre for Supercomputing e.V. (www.gauss-centre.eu). We further thank Peter Behroozi for creating and providing the {\sc rockstar}  halo catalogs and merger trees of the {\sc esmdpl} simulation.

\section*{Data availability} 

The dark matter only simulation, part of the MultiDark project, is available under the name ESMDPL-2048 at https://www.cosmosim.org/cms/files/simulation-data/ . The CosmoSim data base (www.cosmosim.org) is a service by the Leibniz-Institute for Astrophysics Potsdam (AIP). The CoDaII data are available upon reasonable request to the authors.


\bibliographystyle{mnras}

\bibliography{biblicompletenew}

\begin{thebibliography}{56}
\expandafter\ifx\csname natexlab\endcsname\relax\def\natexlab#1{#1}\fi

\bibitem[{{Aubert} {et~al}\mbox{.}(2015){Aubert}, {Deparis}, \&
  {Ocvirk}}]{2015MNRAS.454.1012A}
{Aubert} D., {Deparis} N., {Ocvirk} P., 2015, \mnras, 454, 1012

\bibitem[{{Aubert} {et~al}\mbox{.}(2018){Aubert}, {Deparis}, {Ocvirk},
  {Shapiro}, {Iliev}, {Yepes}, {Gottl{\"o}ber}, {Hoffman}, \&
  {Teyssier}}]{2018ApJ...856L..22A}
{Aubert} D. {et~al.}, 2018, \apjl, 856, L22

\bibitem[{{Aubert} \& {Teyssier}(2008)}]{2008MNRAS.387..295A}
{Aubert} D., {Teyssier} R., 2008, \mnras, 387, 295

\bibitem[{{Barnett} {et~al}\mbox{.}(2017){Barnett}, {Warren}, {Becker},
  {Mortlock}, {Hewett}, {McMahon}, {Simpson}, \&
  {Venemans}}]{2017A&A...601A..16B}
{Barnett} R., {Warren} S.~J., {Becker} G.~D., {Mortlock} D.~J., {Hewett} P.~C.,
  {McMahon} R.~G., {Simpson} C., {Venemans} B.~P., 2017, \aap, 601, A16

\bibitem[{{Bolton} {et~al}\mbox{.}(2011){Bolton}, {Haehnelt}, {Warren},
  {Hewett}, {Mortlock}, {Venemans}, {McMahon}, \&
  {Simpson}}]{2011MNRAS.416L..70B}
{Bolton} J.~S., {Haehnelt} M.~G., {Warren} S.~J., {Hewett} P.~C., {Mortlock}
  D.~J., {Venemans} B.~P., {McMahon} R.~G., {Simpson} C., 2011, \mnras, 416,
  L70

\bibitem[{{Brown} {et~al}\mbox{.}(2014){Brown}, {Tumlinson}, {Geha}, {Simon},
  {Vargas}, {VandenBerg}, {Kirby}, {Kalirai}, {Avila}, {Gennaro}, {Ferguson},
  {Mu{\~n}oz}, {Guhathakurta}, \& {Renzini}}]{2014ApJ...796...91B}
{Brown} T.~M. {et~al.}, 2014, \apj, 796, 91

\bibitem[{{Bullock} {et~al}\mbox{.}(2000){Bullock}, {Kravtsov}, \&
  {Weinberg}}]{2000ApJ...539..517B}
{Bullock} J.~S., {Kravtsov} A.~V., {Weinberg} D.~H., 2000, \apj, 539, 517

\bibitem[{{Busha} {et~al}\mbox{.}(2010){Busha}, {Alvarez}, {Wechsler}, {Abel},
  \& {Strigari}}]{2010ApJ...710..408B}
{Busha} M.~T., {Alvarez} M.~A., {Wechsler} R.~H., {Abel} T., {Strigari} L.~E.,
  2010, \apj, 710, 408

\bibitem[{{Carlesi} {et~al}\mbox{.}(2016){Carlesi}, {Sorce}, {Hoffman},
  {Gottl{\"o}ber}, {Yepes}, {Libeskind}, {Pilipenko}, {Knebe}, {Courtois},
  {Tully}, \& {Steinmetz}}]{2016MNRAS.458..900C}
{Carlesi} E. {et~al.}, 2016, \mnras, 458, 900

\bibitem[{{Christenson} {et~al}\mbox{.}(2021){Christenson}, {Becker},
  {Furlanetto}, {Davies}, {Malkan}, {Zhu}, {Boera}, \&
  {Trapp}}]{2021ApJ...923...87C}
{Christenson} H.~M., {Becker} G.~D., {Furlanetto} S.~R., {Davies} F.~B.,
  {Malkan} M.~A., {Zhu} Y., {Boera} E., {Trapp} A., 2021, \apj, 923, 87

\bibitem[{{Curtis-Lake} {et~al}\mbox{.}(2012){Curtis-Lake}, {McLure}, {Pearce},
  {Dunlop}, {Cirasuolo}, {Stark}, {Almaini}, {Bradshaw}, {Chuter}, {Foucaud},
  \& {Hartley}}]{2012MNRAS.422.1425C}
{Curtis-Lake} E. {et~al.}, 2012, \mnras, 422, 1425

\bibitem[{{Dawoodbhoy} {et~al}\mbox{.}(2018){Dawoodbhoy}, {Shapiro}, {Ocvirk},
  {Aubert}, {Gillet}, {Choi}, {Iliev}, {Teyssier}, {Yepes}, {Gottl{\"o}ber},
  {D'Aloisio}, {Park}, \& {Hoffman}}]{2018MNRAS.480.1740D}
{Dawoodbhoy} T. {et~al.}, 2018, \mnras, 480, 1740

\bibitem[{{Dijkstra} {et~al}\mbox{.}(2014){Dijkstra}, {Wyithe}, {Haiman},
  {Mesinger}, \& {Pentericci}}]{2014MNRAS.440.3309D}
{Dijkstra} M., {Wyithe} S., {Haiman} Z., {Mesinger} A., {Pentericci} L., 2014,
  \mnras, 440, 3309

\bibitem[{{Dubois} \& {Teyssier}(2008)}]{2008A&A...477...79D}
{Dubois} Y., {Teyssier} R., 2008, \aap, 477, 79

\bibitem[{{Elahi} {et~al}\mbox{.}(2016){Elahi}, {Knebe}, {Pearce}, {Power},
  {Yepes}, {Cui}, {Cunnama}, {Kay}, {Sembolini}, {Beck}, {Dav{\'e}},
  {February}, {Huang}, {Katz}, {McCarthy}, {Murante}, {Perret}, {Puchwein},
  {Saro}, \& {Teyssier}}]{2016MNRAS.458.1096E}
{Elahi} P.~J. {et~al.}, 2016, \mnras, 458, 1096

\bibitem[{{Fan} {et~al}\mbox{.}(2006){Fan}, {Strauss}, {Becker}, {White},
  {Gunn}, {Knapp}, {Richards}, {Schneider}, {Brinkmann}, \&
  {Fukugita}}]{2006AJ....132..117F}
{Fan} X. {et~al.}, 2006, \aj, 132, 117

\bibitem[{{Finlator} {et~al}\mbox{.}(2009){Finlator}, {{\"O}zel}, \&
  {Dav{\'e}}}]{2009MNRAS.393.1090F}
{Finlator} K., {{\"O}zel} F., {Dav{\'e}} R., 2009, \mnras, 393, 1090

\bibitem[{{Gnedin}(2000)}]{2000ApJ...542..535G}
{Gnedin} N.~Y., 2000, \apj, 542, 535

\bibitem[{{Gnedin} \& {Abel}(2001)}]{2001NewA....6..437G}
{Gnedin} N.~Y., {Abel} T., 2001, Nature, 6, 437

\bibitem[{{Gunn} \& {Peterson}(1965)}]{1965ApJ...142.1633G}
{Gunn} J.~E., {Peterson} B.~A., 1965, \apj, 142, 1633

\bibitem[{{Iliev} {et~al}\mbox{.}(2014){Iliev}, {Mellema}, {Ahn}, {Shapiro},
  {Mao}, \& {Pen}}]{2014MNRAS.439..725I}
{Iliev} I.~T., {Mellema} G., {Ahn} K., {Shapiro} P.~R., {Mao} Y., {Pen} U.-L.,
  2014, \mnras, 439, 725

\bibitem[{{Iliev} {et~al}\mbox{.}(2006){Iliev}, {Mellema}, {Pen}, {Merz},
  {Shapiro}, \& {Alvarez}}]{2006MNRAS.369.1625I}
{Iliev} I.~T., {Mellema} G., {Pen} U.~L., {Merz} H., {Shapiro} P.~R., {Alvarez}
  M.~A., 2006, \mnras, 369, 1625

\bibitem[{{Iliev} {et~al}\mbox{.}(2005){Iliev}, {Shapiro}, \&
  {Raga}}]{2005MNRAS.361..405I}
{Iliev} I.~T., {Shapiro} P.~R., {Raga} A.~C., 2005, \mnras, 361, 405

\bibitem[{{Kannan} {et~al}\mbox{.}(2021){Kannan}, {Garaldi}, {Smith}, {Pakmor},
  {Springel}, {Vogelsberger}, \& {Hernquist}}]{2021MNRAS.tmp.3440K}
{Kannan} R., {Garaldi} E., {Smith} A., {Pakmor} R., {Springel} V.,
  {Vogelsberger} M., {Hernquist} L., 2021, \mnras

\bibitem[{{Klypin} {et~al}\mbox{.}(1999){Klypin}, {Kravtsov}, {Valenzuela}, \&
  {Prada}}]{1999ApJ...522...82K}
{Klypin} A., {Kravtsov} A.~V., {Valenzuela} O., {Prada} F., 1999, \apj, 522, 82

\bibitem[{{Koposov} {et~al}\mbox{.}(2009){Koposov}, {Yoo}, {Rix}, {Weinberg},
  {Macci{\`o}}, \& {Escud{\'e}}}]{2009ApJ...696.2179K}
{Koposov} S.~E., {Yoo} J., {Rix} H.-W., {Weinberg} D.~H., {Macci{\`o}} A.~V.,
  {Escud{\'e}} J.~M., 2009, \apj, 696, 2179

\bibitem[{{McGreer} {et~al}\mbox{.}(2015){McGreer}, {Mesinger}, \&
  {D'Odorico}}]{2015MNRAS.447..499M}
{McGreer} I.~D., {Mesinger} A., {D'Odorico} V., 2015, \mnras, 447, 499

\bibitem[{{Mu{\~n}oz} {et~al}\mbox{.}(2009){Mu{\~n}oz}, {Madau}, {Loeb}, \&
  {Diemand}}]{2009MNRAS.400.1593M}
{Mu{\~n}oz} J.~A., {Madau} P., {Loeb} A., {Diemand} J., 2009, \mnras, 400, 1593

\bibitem[{{Ocvirk} \& {Aubert}(2011)}]{2011MNRAS.417L..93O}
{Ocvirk} P., {Aubert} D., 2011, \mnras, 417, L93

\bibitem[{{Ocvirk} {et~al}\mbox{.}(2013){Ocvirk}, {Aubert}, {Chardin}, {Knebe},
  {Libeskind}, {Gottl{\"o}ber}, {Yepes}, \& {Hoffman}}]{2013ApJ...777...51O}
{Ocvirk} P., {Aubert} D., {Chardin} J., {Knebe} A., {Libeskind} N.,
  {Gottl{\"o}ber} S., {Yepes} G., {Hoffman} Y., 2013, \apj, 777, 51

\bibitem[{{Ocvirk} {et~al}\mbox{.}(2020){Ocvirk}, {Aubert}, {Sorce}, {Shapiro},
  {Deparis}, {Dawoodbhoy}, {Lewis}, {Teyssier}, {Yepes}, {Gottl{\"o}ber},
  {Ahn}, {Iliev}, \& {Hoffman}}]{2020MNRAS.tmp.1438O}
{Ocvirk} P. {et~al.}, 2020, \mnras

\bibitem[{{Ocvirk} {et~al}\mbox{.}(2014){Ocvirk}, {Gillet}, {Aubert}, {Knebe},
  {Libeskind}, {Chardin}, {Gottl{\"o}ber}, {Yepes}, \&
  {Hoffman}}]{2014ApJ...794...20O}
{Ocvirk} P. {et~al.}, 2014, \apj, 794, 20

\bibitem[{{Ocvirk} {et~al}\mbox{.}(2016){Ocvirk}, {Gillet}, {Shapiro},
  {Aubert}, {Iliev}, {Teyssier}, {Yepes}, {Choi}, {Sullivan}, {Knebe},
  {Gottl{\"o}ber}, {D'Aloisio}, {Park}, {Hoffman}, \&
  {Stranex}}]{2016MNRAS.463.1462O}
{Ocvirk} P. {et~al.}, 2016, \mnras, 463, 1462

\bibitem[{{Pentericci} {et~al}\mbox{.}(2011){Pentericci}, {Fontana},
  {Vanzella}, {Castellano}, {Grazian}, {Dijkstra}, {Boutsia}, {Cristiani},
  {Dickinson}, {Giallongo}, {Giavalisco}, {Maiolino}, {Moorwood}, {Paris}, \&
  {Santini}}]{2011ApJ...743..132P}
{Pentericci} L. {et~al.}, 2011, \apj, 743, 132

\bibitem[{{Pentericci} {et~al}\mbox{.}(2014){Pentericci}, {Vanzella},
  {Fontana}, {Castellano}, {Treu}, {Mesinger}, {Dijkstra}, {Grazian}, {Brada{\v
  c}}, {Conselice}, {Cristiani}, {Dunlop}, {Galametz}, {Giavalisco},
  {Giallongo}, {Koekemoer}, {McLure}, {Maiolino}, {Paris}, \&
  {Santini}}]{2014ApJ...793..113P}
{Pentericci} L. {et~al.}, 2014, \apj, 793, 113

\bibitem[{{Petkova} \& {Springel}(2009)}]{2009MNRAS.396.1383P}
{Petkova} M., {Springel} V., 2009, \mnras, 396, 1383

\bibitem[{{Planck Collaboration} {et~al}\mbox{.}(2014){Planck Collaboration},
  {Ade}, {Aghanim}, {Armitage-Caplan}, {Arnaud}, {Ashdown}, {Atrio-Barandela},
  {Aumont}, {Baccigalupi}, {Banday}, \& et~al.}]{2014A&A...571A..16P}
{Planck Collaboration} {et~al.}, 2014, \aap, 571, A16

\bibitem[{{Planck Collaboration} {et~al}\mbox{.}(2016){Planck Collaboration},
  {Ade}, {Aghanim}, {Arnaud}, {Ashdown}, {Aumont}, {Baccigalupi}, {Banday},
  {Barreiro}, {Bartlett}, \& et~al.}]{2016A&A...594A..13P}
{Planck Collaboration} {et~al.}, 2016, \aap, 594, A13

\bibitem[{{Rosdahl} {et~al}\mbox{.}(2013){Rosdahl}, {Blaizot}, {Aubert},
  {Stranex}, \& {Teyssier}}]{2013MNRAS.436.2188R}
{Rosdahl} J., {Blaizot} J., {Aubert} D., {Stranex} T., {Teyssier} R., 2013,
  \mnras, 436, 2188

\bibitem[{{Schenker} {et~al}\mbox{.}(2012){Schenker}, {Stark}, {Ellis},
  {Robertson}, {Dunlop}, {McLure}, {Kneib}, \& {Richard}}]{2012ApJ...744..179S}
{Schenker} M.~A., {Stark} D.~P., {Ellis} R.~S., {Robertson} B.~E., {Dunlop}
  J.~S., {McLure} R.~J., {Kneib} J.-P., {Richard} J., 2012, \apj, 744, 179

\bibitem[{{Shapiro} {et~al}\mbox{.}(1994){Shapiro}, {Giroux}, \&
  {Babul}}]{1994ApJ...427...25S}
{Shapiro} P.~R., {Giroux} M.~L., {Babul} A., 1994, \apj, 427, 25

\bibitem[{{Shapiro} {et~al}\mbox{.}(2004){Shapiro}, {Iliev}, \&
  {Raga}}]{2004MNRAS.348..753S}
{Shapiro} P.~R., {Iliev} I.~T., {Raga} A.~C., 2004, \mnras, 348, 753

\bibitem[{{Sorce}(2015)}]{2015MNRAS.450.2644S}
{Sorce} J.~G., 2015, \mnras, 450, 2644

\bibitem[{{Sorce}(2018)}]{2018MNRAS.478.5199S}
{Sorce} J.~G., 2018, \mnras, 478, 5199

\bibitem[{{Sorce} {et~al}\mbox{.}(2019){Sorce}, {Blaizot}, \&
  {Dubois}}]{2019MNRAS.486.3951S}
{Sorce} J.~G., {Blaizot} J., {Dubois} Y., 2019, \mnras, 486, 3951

\bibitem[{{Sorce} {et~al}\mbox{.}(2016{\natexlab{a}}){Sorce}, {Gottl{\"o}ber},
  {Hoffman}, \& {Yepes}}]{2016MNRAS.460.2015S}
{Sorce} J.~G., {Gottl{\"o}ber} S., {Hoffman} Y., {Yepes} G.,
  2016{\natexlab{a}}, \mnras, 460, 2015

\bibitem[{{Sorce} {et~al}\mbox{.}(2016{\natexlab{b}}){Sorce}, {Gottl{\"o}ber},
  {Yepes}, {Hoffman}, {Courtois}, {Steinmetz}, {Tully}, {Pomar{\`e}de}, \&
  {Carlesi}}]{2016MNRAS.455.2078S}
{Sorce} J.~G. {et~al.}, 2016{\natexlab{b}}, \mnras, 455, 2078

\bibitem[{{Springel}(2005)}]{Springel2005}
{Springel} V., 2005, \mnras, 364, 1105

\bibitem[{{Stark} {et~al}\mbox{.}(2010){Stark}, {Ellis}, {Chiu}, {Ouchi}, \&
  {Bunker}}]{2010MNRAS.408.1628S}
{Stark} D.~P., {Ellis} R.~S., {Chiu} K., {Ouchi} M., {Bunker} A., 2010, \mnras,
  408, 1628

\bibitem[{{Teyssier}(2002)}]{2002A&A...385..337T}
{Teyssier} R., 2002, \aap, 385, 337

\bibitem[{{Th{\'e}lie} {et~al}\mbox{.}(2021){Th{\'e}lie}, {Aubert}, {Gillet},
  \& {Ocvirk}}]{2021arXiv211111910T}
{Th{\'e}lie} E., {Aubert} D., {Gillet} N., {Ocvirk} P., 2021, arXiv e-prints,
  arXiv:2111.11910

\bibitem[{{Tilvi} {et~al}\mbox{.}(2014){Tilvi}, {Papovich}, {Finkelstein},
  {Long}, {Song}, {Dickinson}, {Ferguson}, {Koekemoer}, {Giavalisco}, \&
  {Mobasher}}]{2014ApJ...794....5T}
{Tilvi} V. {et~al.}, 2014, \apj, 794, 5

\bibitem[{{Trac} \& {Cen}(2007)}]{2007ApJ...671....1T}
{Trac} H., {Cen} R., 2007, \apj, 671, 1

\bibitem[{{Trac} \& {Gnedin}(2011)}]{2011ASL.....4..228T}
{Trac} H.~Y., {Gnedin} N.~Y., 2011, Advanced Science Letters, 4, 228

\bibitem[{{Treu} {et~al}\mbox{.}(2013){Treu}, {Schmidt}, {Trenti}, {Bradley},
  \& {Stiavelli}}]{2013ApJ...775L..29T}
{Treu} T., {Schmidt} K.~B., {Trenti} M., {Bradley} L.~D., {Stiavelli} M., 2013,
  \apjl, 775, L29

\bibitem[{{Weinmann} {et~al}\mbox{.}(2007){Weinmann}, {Macci{\`o}}, {Iliev},
  {Mellema}, \& {Moore}}]{2007MNRAS.381..367W}
{Weinmann} S.~M., {Macci{\`o}} A.~V., {Iliev} I.~T., {Mellema} G., {Moore} B.,
  2007, \mnras, 381, 367

\end{thebibliography}
 \label{lastpage}
\end{document}